\documentclass[twocolumn,showpacs,superscriptaddress]{revtex4-1}
\usepackage{graphicx}
\usepackage{units}
\usepackage{multirow}
\usepackage{bigstrut}
\usepackage{array}
\newcommand{\jpsi}{J/\psi}
\newcommand{\chicj}{\chi_{cJ}}
\newcommand{\chiczero}{\chi_{c0}}
\newcommand{\chicone}{\chi_{c1}}
\newcommand{\chictwo}{\chi_{c2}}
\newcommand{\psip}{\psi(3686)}
\newcommand{\gm}{\gamma}
\newcommand{\gmsm}{\gamma_{\unit{sm}}}
\newcommand{\gmlg}{\gamma_{\unit{lg}}}
\newcommand{\rmgmsm}{M_{\gmsm-\unit{recoil}}}
\newcommand{\rmgg}{M_{\gm\gm-\unit{recoil}}}
\newcommand{\dil}{\ell^+\ell^-}
\newcommand{\ee}{e^+e^-}
\newcommand{\mm}{\mu^+\mu^-}
\newcommand{\mev}{\,\unit{MeV}}
\newcommand{\mevcc}{\,\unit{MeV}/c^2}
\newcommand{\gev}{\,\unit{GeV}}
\newcommand{\gevc}{\,\unit{GeV}/c}
\newcommand{\gevcc}{\,\unit{GeV}/c^2}
\newcommand{\br}[1]{\mathcal{B}(#1)}
\newcommand{\chisq}[1]{\chi^{2}_{\unit{#1}}}
\newcommand {\eg}       {\emph{e}.\emph{g}.}
\newcommand {\ie}       {\emph{i}.\emph{e}.}

\begin{document}

\title{\boldmath Evidence for the Direct Two-Photon Transition from $\psip$ to $\jpsi$}

\author{
\small
\begin{center}
M.~Ablikim$^{1}$, M.~N.~Achasov$^{5}$, D.~J.~Ambrose$^{40}$, F.~F.~An$^{1}$, Q.~An$^{41}$, Z.~H.~An$^{1}$, J.~Z.~Bai$^{1}$, R.~B.~Ferroli$^{18}$, Y.~Ban$^{27}$, J.~Becker$^{2}$
% ...
\\
\vspace{0.2cm}
(BESIII Collaboration)\\
\vspace{0.2cm} {\it
$^{1}$ Institute of High Energy Physics, Beijing 100049, P. R. China\\
$^{2}$ Bochum Ruhr-University, 44780 Bochum, Germany\\
% ...
$^{5}$ G.I. Budker Institute of Nuclear Physics SB RAS (BINP), Novosibirsk 630090, Russia\\
% ...
$^{18}$ INFN Laboratori Nazionali di Frascati , Frascati, Italy\\
% ...
$^{27}$ Peking University, Beijing 100871, P. R. China\\
% ...
$^{40}$ University of Rochester, Rochester, New York 14627, USA\\
$^{41}$ University of Science and Technology of China, Hefei 230026, P. R. China\\
% ...
}\end{center}
\vspace{0.4cm}}

\begin{abstract}

The two-photon transition $\psip\to\gm\gm\jpsi$ is studied in a
sample of 106 million $\psip$ decays collected by the BESIII
detector. The branching fraction is measured to be
$(3.1\pm0.6(\unit{stat})^{+0.8}_{-1.0}(\unit{syst}))
\times10^{-4}$ using $\jpsi\to\ee$ and
$\jpsi\to\mm$ decays, and its upper limit is estimated to be $4.5\times10^{-4}$ at the 90\% conference level. This work represents the first
measurement of a two-photon transition among charmonium states. The orientation of the $\psip$ decay plane and the $\jpsi$
polarization in this decay are also studied. In addition, the product branching fractions of sequential $E1$ transitions $\psip\to\gm\chicj, \chicj\to\gm\jpsi (J=0,1,2)$ are reported.

\end{abstract}


\maketitle


The XYZ~\cite{Brambilla:2010cs} particles, which do not
fit potential model expectations in QCD theory, have been a key
challenge to the QCD description of charmonium-like
states~\cite{Asner:2008nq}. To fully understand those states, it is
necessary to consider the coupling of a charmonium state to a
$D\bar{D}$ meson pair. These coupled-channel effects,
which also play an important role in the charmonium transitions of
low lying states (\ie, from $\psip$ to $\jpsi$), are especially
relevant for the radiative transition
processes~\cite{Eichten:2004uh}. In the well-known electric dipole
transitions, the strength of coupled-channel effects will likely be
hard to establish, since the accompanying relativistic corrections
may be more important~\cite{Li:2009zu}. However, the two-photon
transition $\psip\to\gm\gm\jpsi$ is more sensitive to the
coupled-channel effect and thus provides a unique opportunity to
investigate these issues~\cite{He:2010pb}.

Two-photon spectroscopy has been a very powerful tool for the study
of the excitation spectra of a variety of systems with a wide range of sizes,
such as molecules, atomic hydrogen and positronium~\cite{Pachucki:1996jw}.
Studying the analogous process in quarkonium states is a natural extension
of this work, in order to gain insight into non-perturbative QCD phenomena.
But so far, two-photon transitions in quarkonia have
eluded experimental observation~\cite{Bai:2004cg,Adam:2005uh,:2008kb}.
For example, in a study of $\psip\to\gm\chi_{cJ}(J=0,1,2)$ reported by
CLEO-c~\cite{:2008kb}, the upper limit for $\br{\psip\to\gm\gm\jpsi}$ was
estimated to be $1\times10^{-3}$.

This Letter presents the first evidence for the two-photon transition
$\psip\to\gm\gm\jpsi$, as well as studies of the orientation
of the $\psip$ decay plane and the $\jpsi$ polarization in the decay.
The branching fractions of double $E1$ transitions $\psip\to\gm(\gm\jpsi)_{\chicj}$ through $\chicj$ intermediate states are also reported. The data analyzed were obtained by the BESIII experiment~\cite{:2009vd}
viewing electron-positron collisions  at the BEPCII collider.
An integrated luminosity of 156.4 $\rm{pb}^{-1}$ was obtained at a
center-of-mass energy $\sqrt{s} = M(\psip) = 3.686\gev$.
The number of $\psip$ decays in this sample is estimated to be
$(1.06\pm0.04)\times10^8$~\cite{Ablikim:2010zn}.  In addition,
42.6 $\rm{pb}^{-1}$ of continuum data were taken below the $\psip$, at $\sqrt{s}=3.65\gev$,
to evaluate the potential backgrounds from non-resonant events.


The upgraded BEPCII~\cite{Zhang:2010zz} at Beijing is a two-ring
electron-positron collider.  The BESIII detector~\cite{:2009vd}
is an approximately cylindrically symmetric detector which covers
$93\%$ of the solid angle around the collision point.
In order of increasing distance from the interaction point,
the sub-detectors include a 43-layer
main wire drift chamber (MDC), a time-of-flight (TOF) system with two
layers in the barrel region and one layer for each end-cap,
and a 6240 cell CsI(Tl) crystal electro-magnetic calorimeter (EMC)
with both barrel and endcap sections.
The barrel components reside within a superconducting solenoid
magnet providing a 1.0\,T magnetic field aligned with the beam axis.
Finally, there is a muon chamber consisting of nine layers of resistive plate
chambers within the return yoke of the magnet.
The momentum resolution for charged tracks in the MDC
is $0.5\%$ for transverse momenta of $1\gevc$.
The energy resolution for showers in the EMC is $2.5\%$ for $1\gev$ photons.

\begin{figure}[t]

\begin{minipage}[t]{\linewidth}
\begin{flushleft}
\hspace{0.02cm}
\includegraphics[width=0.6\linewidth]{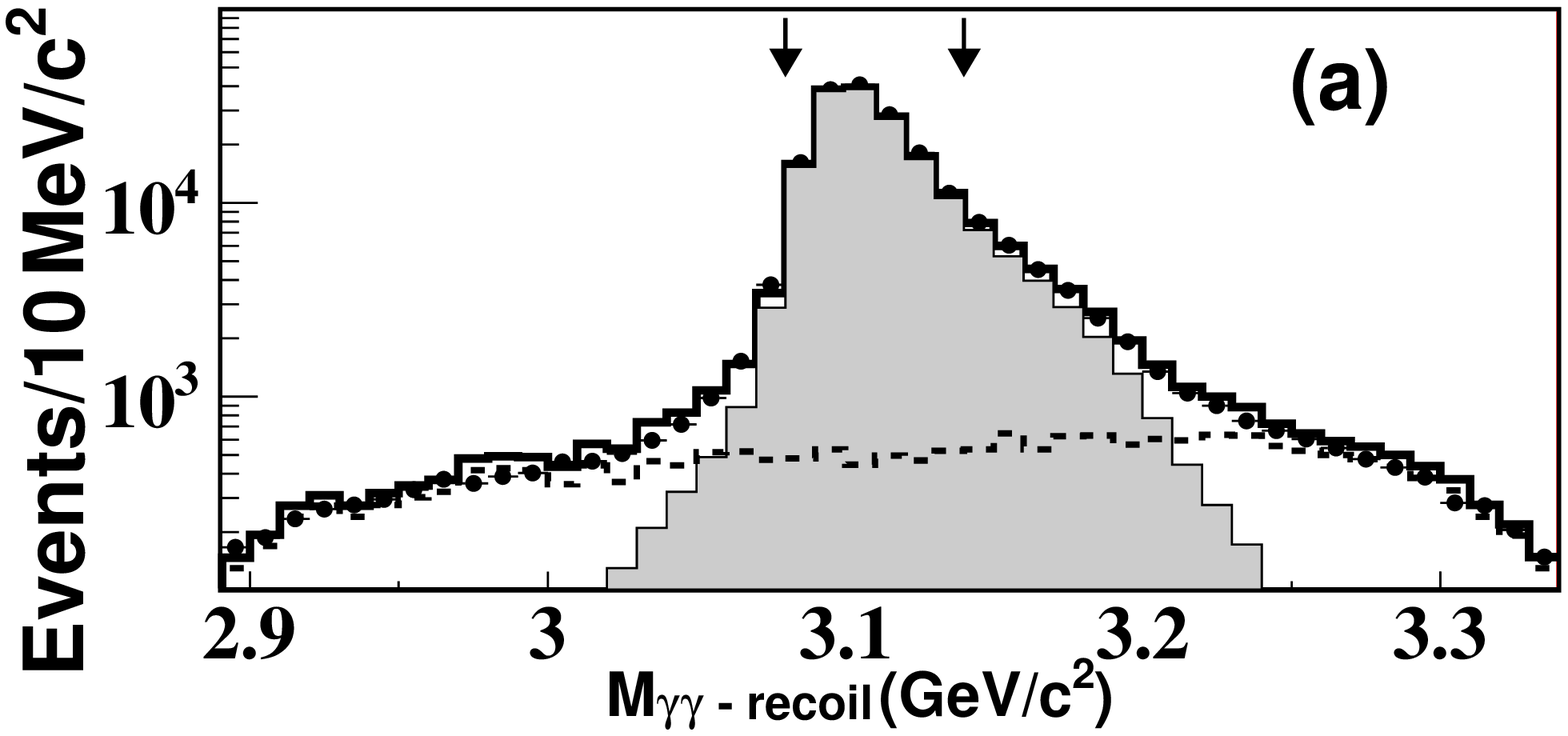}
\end{flushleft}
\end{minipage}

\vspace{0.05cm}

\begin{minipage}[t]{\linewidth}
\begin{flushleft}
\includegraphics[width=\linewidth]{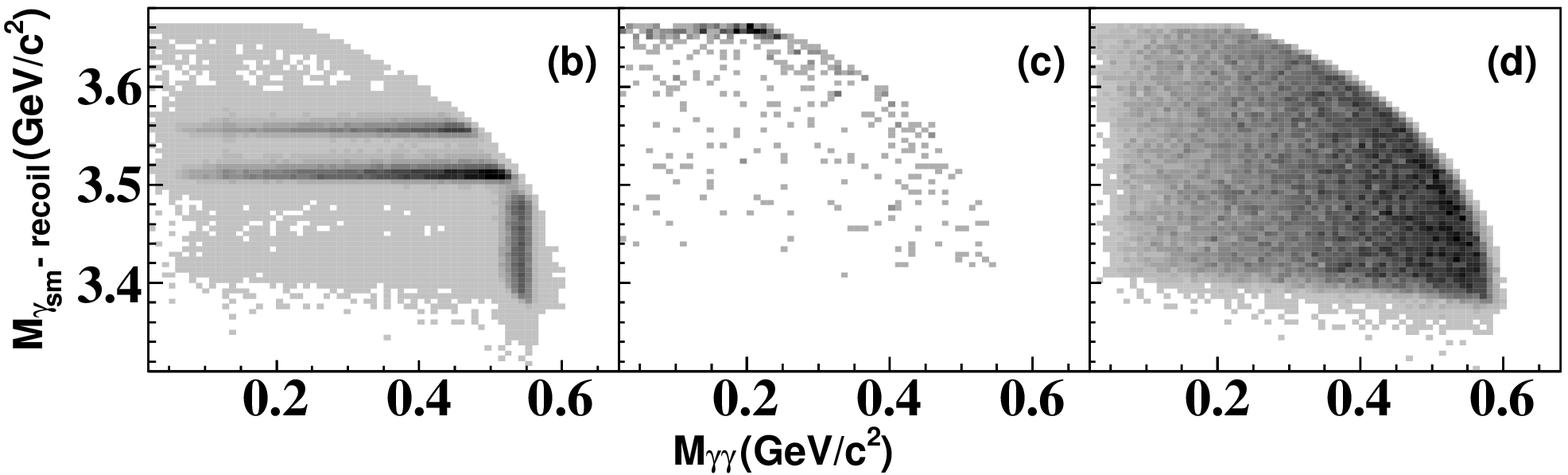}
\end{flushleft}
\end{minipage}
 \vspace{-0.75cm}
 \caption{Up (a): distributions of $\rmgg$ in data (points) and in the combined dataset (solid line) of MC simulation of $\psip$ decays (shaded histogram) and continuum backgrounds (dashed line), before the KF is applied. The arrows indicate the window to select a $\jpsi$ candidate; Down: scatter plots of $\rmgmsm$ versus $M_{\gm\gm}$ for the $\gm\gm\ee$ channel, in data (b), continuum data (c), MC simulated signal (d), after applying the KF constrain and the $\rmgg$ window.
 The corresponding plots for the $\gm\gm\mm$ channel are very similar.}
\vspace{-0.5cm}
\label{fig:scat}
\end{figure}

This work studies $\psip\to\gm\gm\jpsi$ followed by $\jpsi\to\dil$
($\ell$ denotes $e$ or $\mu$), which is referred to as the signal process.
Events selected contain exactly two oppositely charged good tracks in
the MDC tracking system, corresponding to the dilepton from $\jpsi$ decay.
The requirements to judge a track as good include $|\cos\theta| < 0.93$
($\theta$ is the polar angle with respect to the beam direction),
and the minimum distance of approach between the track and the production
vertex less than 10\,cm along the beam axis
and less than 1\,cm projected in the perpendicular plane.
The lepton is identified with the
ratio of EMC shower energy to MDC track momentum,
$E/p$, which must be larger than 0.7 for
an electron, or smaller than 0.6 for a muon.
To suppress non-$\jpsi$ decay leptons,
we require the momentum of each lepton to be larger than $0.8\gevc$.
A vertex fit (VF) constrains the production vertex, \
which is updated run-by-run,
and the tracks of the dilepton candidates to a common vertex;
only events with $\chisq{VF}/\unit{d.o.f.}<20$ are accepted.

Reconstructed EMC showers unmatched to either charged track
and with an energy larger than $25\mev$ in the barrel region ($|\cos\theta|<0.80$)
or larger than $50\mev$ in the end-caps ($0.86<|\cos\theta|<0.92$)
are used as photon candidates.
To reject bremsstrahlung photons, showers matching the initial momentum
of either lepton within $10^\circ$ are also discarded.
Showers from noise, not originating from the beam collision,
are suppressed by requiring the EMC cluster time to lie within a 700\,ns
window near the event start time.

Events are required to have only two photon candidates.
A kinematic fit (KF) constrains the
vertexed dilepton to the nominal mass of the intermediate $\jpsi$, and the resulting $\jpsi$ and photon candidates to the known initial four-momentum of the $\psip$.
The KF fit quality
$\chisq{KF}$ is required to be $\chisq{KF}/\unit{d.o.f.}<12$. For convenience, we
use $\gmlg$ ($\gmsm$) to denote the larger (smaller) energy photon.
As indicated in Fig.~\ref{fig:scat}(a), $\jpsi$ candidates are identified with the requirement that the recoil mass of the two photons, $\rmgg$, is within $(3.08, 3.14)\gevcc$.

Scatter plots of recoiling mass $\rmgmsm$ from the lower energy
photon $\gmsm$ versus invariant mass of two photons $M_{\gm\gm}$ are
shown in Fig.~\ref{fig:scat}, where clear resonance bands are seen
from the decays $\psip\to\gm\chicj(J=0,1,2)$ (three horizontal
bands) and $\psip\to\pi^0(\eta)\jpsi$ (two vertical bands).
As indicated in Fig.~\ref{fig:scat}(c), the continuum backgrounds are most dominant at the top of the plots, of which the primary sources include the bhabha scattering, the dimuon process and the ISR production of $\jpsi$.
These backgrounds are excluded by discarding events with $\rmgmsm>3.6\gevcc$.
To suppress backgrounds from $\psip\rightarrow\pi^0(\eta)J/\psi$, the diphoton invariant mass
$M_{\gamma\gamma}$ is required to be larger than $0.15\gevcc$ and the recoil momentum of the
diphoton must be larger than $0.25\gevc$.

\begin{figure}[tp!]
\centering
\includegraphics[width=0.95\linewidth]{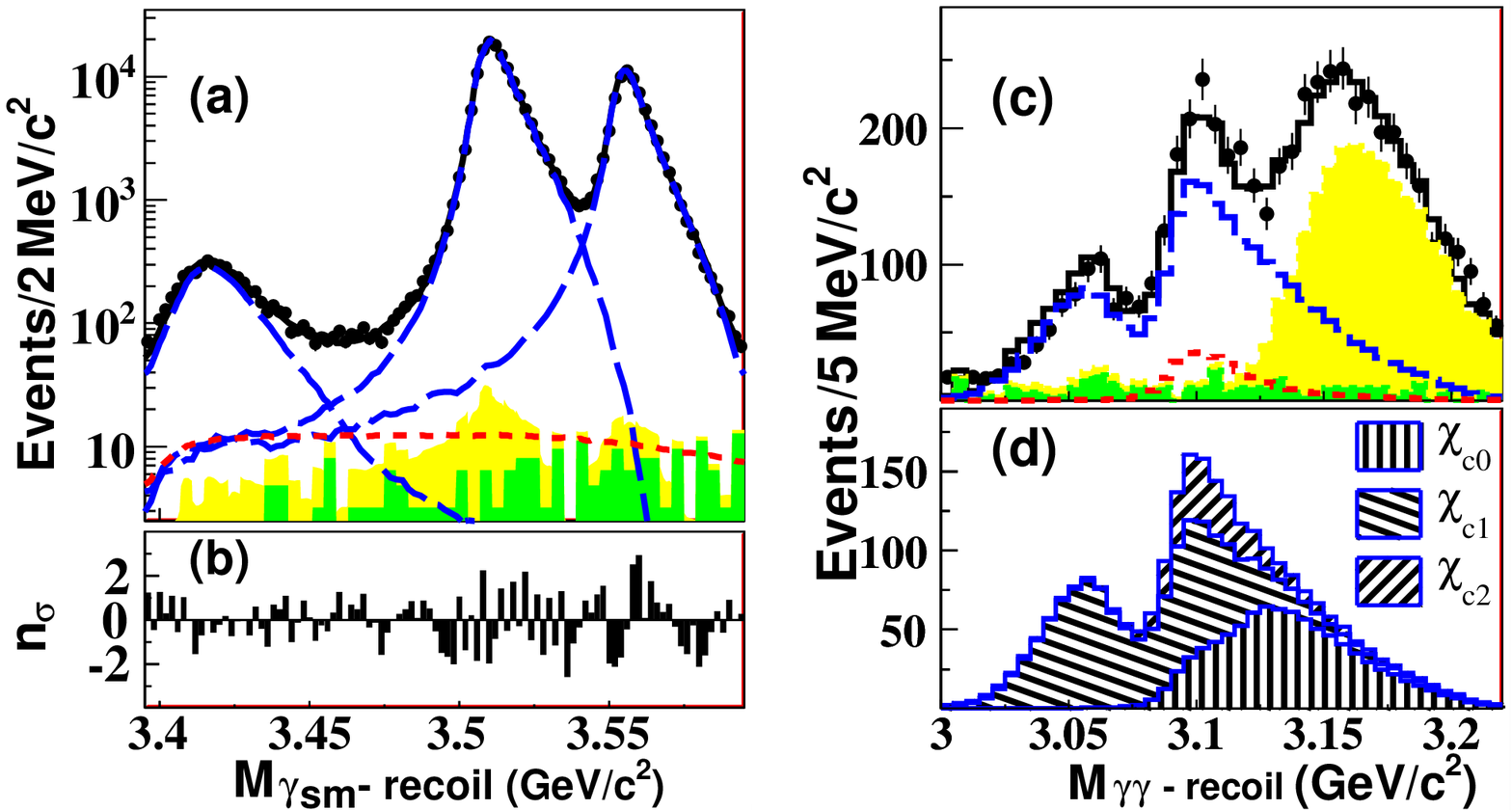}
\caption{(color online) Plot a: unbinned maximum likelihood fit to the
   distribution of $\rmgmsm$ in data with combination of the two $\jpsi$-decay modes.
   Thick lines are the sum of the fitting models and long-dashed
   lines are the $\chicj$ shapes. Short-dashed lines represent
   the two-photon signal processes. Shaded histograms are $\psip$-decay backgrounds (yellow) and non-$\psip$ backgrounds (green), with the fixed amplitude and shape taken from MC simulation and continuum data. Plot b: the number of standard deviations, $n_\sigma$, of data points from the fitted curves in plot a. The rates of the signal process and sequential $\chicj$ processes are derived from these fits.
   Plot c: distributions of $\rmgg$ in data (signals and known backgrounds) with the kinematic requirement $3.44\gevcc<\rmgmsm<3.48\gevcc$ and with the removal of $\chisq{KF}$ and $\rmgg$ restrictions.
   Plot d: stacked histograms of the three $\chicj$ components in plot c. }
\vspace{-0.5cm}
\label{fig:global_fit}
\end{figure}

Monte Carlo (MC) simulations of $\psip$ decays are used to understand
the backgrounds and also to estimate the detection efficiency.
At BESIII, the simulation includes the beam energy spread and treats
the initial-state radiation with KKMC~\cite{Jadach:2000ir}.
Specific decay modes from the PDG~\cite{pdg2012}
are modeled with EVTGEN~\cite{ref:bes3gen},
and the unknown decay modes with Lundcharm~\cite{Chen:2000tv}.
The detector response is described using GEANT4~\cite{ref:geant4}.
For the $\psip\to\gm\gm\jpsi$ channel, the momenta of
decay particles are simulated according to the measured polarization structure in this work.
Generic $\psip$ decay samples serve for understanding the background channels;
dominant backgrounds were generated with high statistics.
Angular distributions of the cascade $E1$ transitions
$\psip\to\gm\chicj\to\gm\gm\jpsi$
are assumed to follow the formulae in Ref.~\cite{Karl:1975jp}.
Note that the $\chicj$ line shapes were simulated with the Breit-Wigner
distributions weighted with $E^3_{\gamma^*_1}E^3_{\gamma^*_2}$ to account for the double $E1$ transitions,
and extended out to $\pm 200\mevcc$ away from the nominal masses,
using masses and widths in PDG~\cite{pdg2012}.
Here, $E_{\gamma^*_1}$($E_{\gamma^*_2}$) is the energy of the radiative photon $\gamma^*_1$($\gamma^*_2$) in the rest frame of
the mother particle $\psip$($\chicj$).

The yield of the signal process $\psip\to\gm\gm\jpsi$, together with those of the cascade $E1$ transition processes, is estimated by a global fit to
the spectrum of $\rmgmsm$. The fit results are shown in Fig.~\ref{fig:global_fit}.
The shape and magnitude of $\psip$-decay backgrounds were fixed based on MC simulation.
Non-$\psip$ decay backgrounds are estimated in continuum data, scaling by luminosity and the $1/s$ dependence of the cross sections. This scaling is verified by the good description of the $\jpsi$ backgrounds in the $\rmgg$ distribution shown in Fig.~\ref{fig:scat}(a).
The distributions of the signal process and the cascade $E1$ process are taken from the reconstructed shapes in MC simulation of the modes and smeared with an asymmetric Gaussian with free parameters, which is used to compensate for the difference in line shape between MC and data. By taking the MC shape, detector resolution and wrong assignment of the $E1$ photon are taken into account.
The quality of goodness-of-fit test, $\chi^2/$d.o.f.$=108.0/94=1.15$
in the $\gm\gm\ee$ mode and  $124.8/94=1.33$ in the $\gm\gm\mm$ mode.
The observed signal yields are given in Table~\ref{tab:res}.
The $\psip\to\gm\gm\jpsi$ transition is observed with a statistical significance of 6.6$\sigma$, as determined by the ratio of the maximum likelihood value and the likelihood value for a fit with null-signal hypothesis. When the systematic uncertainties are taken into account with the assumption of Gaussian distributions, the significance is evaluated to be 3.8$\sigma$, which
corresponds to a probability of a background fluctuation to the observed signal yield of $7.2\times10^{-5}$. The upper limit for $\br{\psip\to\gm\gm\jpsi}$ is estimated to be $4.5\times10^{-4}$ at the 90\% confidence level including systematic uncertainties.

In calculating $\br{\gm\gm\jpsi}$, a correction factor is included due to the interferences among $\chicj$ states. This effect was checked by the variations of the observed signals in the global fit with inclusion of a floating interference component, which is modeled by the detector-smeared shape of a theoretical calculation~\cite{He:2010pb}. It is found that relative changes on the signal yields are negative with lower bound of $-10\%$. Hence, a correction factor $0.95$ is assigned and $5\%$ is taken as systematic uncertainty.

A cross-check on our procedures is performed with the $\rmgg$ spectrum for the events in the region $3.44\gevcc<\rmgmsm<3.48\gevcc$ without restrictions on $\chisq{KF}$ and $\rmgg$, as shown in Fig.~\ref{fig:global_fit}(c). An excess of data above known backgrounds can be seen around the $\jpsi$ nominal mass, which is expected from the sought-after two-photon process. With the inclusion of the estimated yields of the signal process, the excess is well understood. The high-mass peak above the $\jpsi$ peak comes from the backgrounds of $\psip\to\pi^0\pi^0\jpsi$ decays. This satellite peak can be well described in MC simulation. In Fig.~\ref{fig:global_fit}(d), the three $\chicj$ tails show distinguishable distributions; the small left bump is from the
$\chicone$ tail, while the $\chiczero$ tail is dominant at the right side. The distribution in data in Fig.~\ref{fig:global_fit}(c) can only be well described by the simulated $\chicj$ shapes.

\begin{table}[tp!]
  \vspace{-0.25cm}
  \scriptsize
  \caption{\footnotesize For different channels: the number of observed signals
  $n_e$ ($n_\mu$) and detection efficiency $\epsilon_e$ ($\epsilon_\mu$)
  in $\gm\gm\ee$ ($\gm\gm\mm$) mode; the absolute branching fractions.
  On the bottom, the relative branching fractions
  $R_{MN}\equiv\mathcal{B}_{\chi_{cM}}/\mathcal{B}_{\chi_{cN}}$,
  where $\mathcal{B}_{\chicj}\equiv\br{\psip\to\gm(\gm\jpsi)_{\chicj}}$ are
  listed. Here the first errors are statistical and the second are systematic.}
  \label{tab:res}
  \vspace{-0.3cm}
  \setlength{\extrarowheight}{1.pt}
  \begin{center}
    \begin{tabular}{lccccc} \hline \hline
    {\footnotesize Channels}  & {\footnotesize $n_e$} & {\footnotesize $\epsilon_e$(\%)} & {\footnotesize $n_\mu$} & {\footnotesize $\epsilon_\mu$(\%)}
    & {\footnotesize $\mathcal{B}$($\times 10^{-4}$)}  \\

    $\gm\gm\jpsi$  & 564$\pm$116 & 22.4 & 536$\pm$128 & 30.0
     &   $3.1\pm0.6^{+0.8}_{-1.0}$ \\
    $\gm(\gm\jpsi)_{\chiczero}$  & 1801$\pm$60  & 19.3 & 2491$\pm$69 & 26.0
     &  $15.1\pm0.3\pm1.0$  \\
    $\gm(\gm\jpsi)_{\chicone}$ & 59953$\pm$253 & 28.5 & 81922$\pm$295 & 38.2
     &  $337.7\pm0.9\pm18.3$  \\
     $\gm(\gm\jpsi)_{\chictwo}$  & 32171$\pm$187 & 27.5 & 44136$\pm$219 & 37.1
     &  $187.4\pm0.7\pm10.2$ \\ \hline

    \multicolumn{2}{c}{\footnotesize $R_{21}\equiv\frac{\mathcal{B}_{\chictwo}}{\mathcal{B}_{\chicone}}(\%)$}  & \multicolumn{2}{c}{\footnotesize $R_{01}\equiv\frac{\mathcal{B}_{\chiczero}}{\mathcal{B}_{\chicone}}(\%)$} & \multicolumn{2}{c}{\footnotesize $R_{02}\equiv\frac{\mathcal{B}_{\chiczero}}{\mathcal{B}_{\chictwo}}(\%)$} \\

    \multicolumn{2}{c}{$55.47\pm0.26\pm0.11$} &
    \multicolumn{2}{c}{$4.45\pm0.09\pm0.18$}  &
    \multicolumn{2}{c}{$8.03\pm0.17\pm0.33$} \\ \hline \hline
   \end{tabular}
  \end{center}
 \vspace{-0.5cm}
\end{table}

\begin{figure}[t]
   \centering
   \footnotesize
   \includegraphics[width=0.36\linewidth]{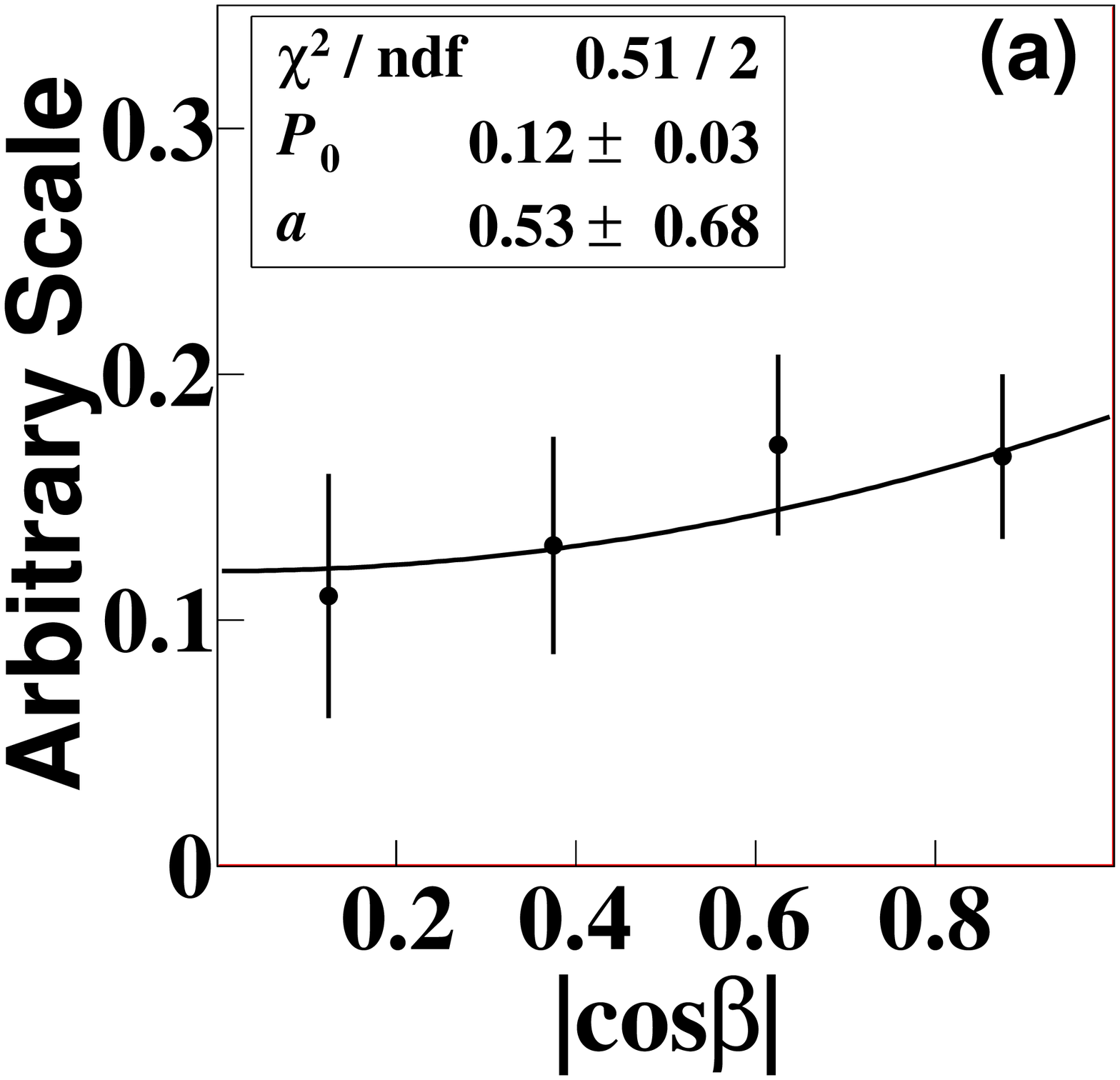}
   \hspace{20pt}
   \includegraphics[width=0.36\linewidth]{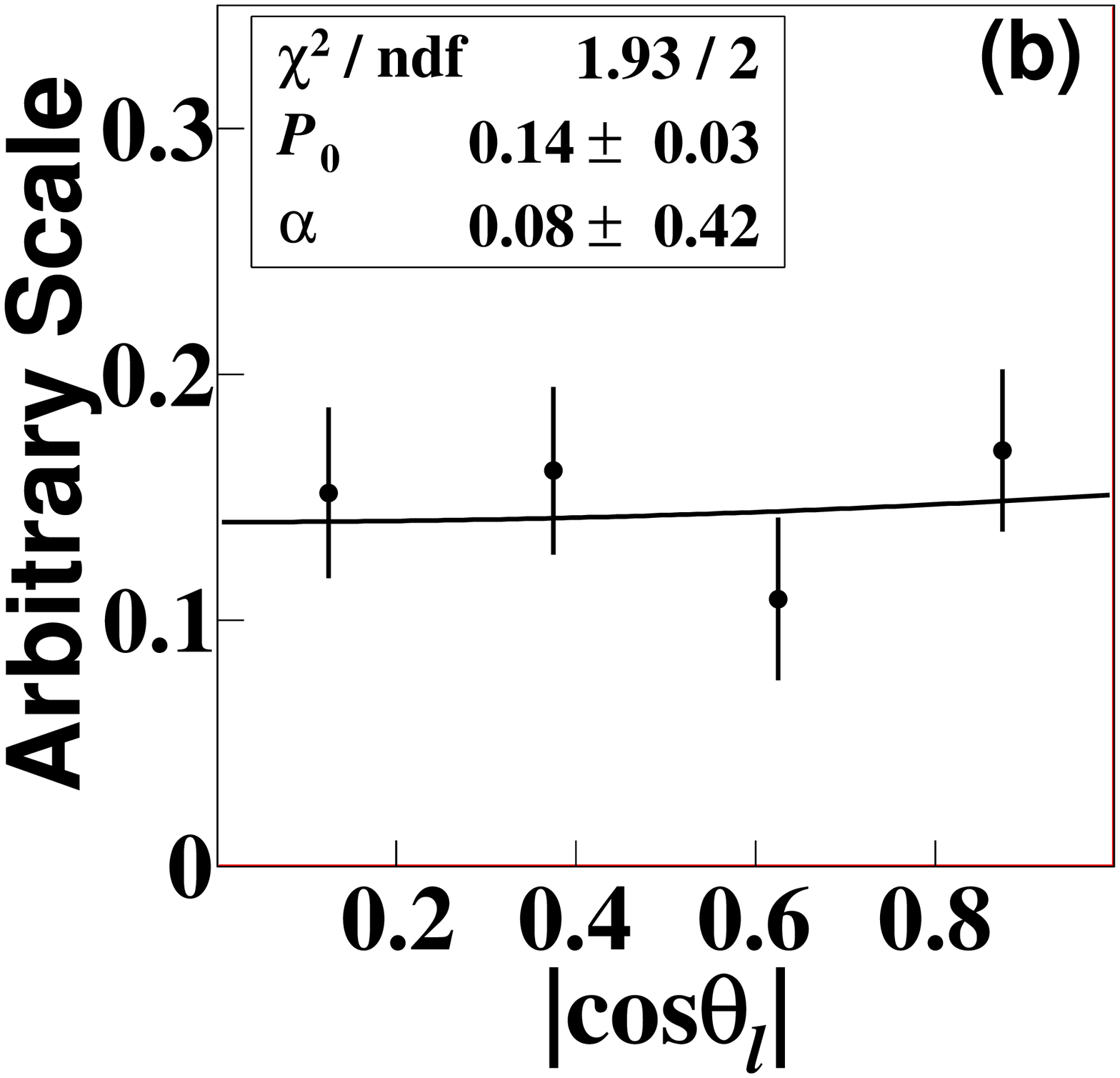}
   \vspace{-0.5cm}
   \caption{(a) The corrected distribution of
   the normal angle $\beta$ of the $\psip$ decay plane, and
   (b) the helicity angle $\theta_\ell$ of $\jpsi$ decays.
   The curves in (a) and (b) present the fits of
   functions $P_0(1+a\cos^2\beta)$ and
   $P_0(1+\alpha\cos^2\theta_\ell)$, respectively.}
   \vspace{-0.35cm}
   \label{fig:decay_fit}
\end{figure}

The angle of the normal axis of the $\psip$ decay plane with
respect to the $\psip$ polarization vector (aligned to the
beam axis), $\beta$, can be determined in our data.
The event rate may be expressed, to leading order, as
$\frac{d\,N}{d\,\cos\beta} \propto 1+ a \cos^2\beta$.
The measurement was carried out in the rest frame of the $\psip$
and the decay plane of the $\psip$ was determined with the momenta
of the two decay particles $\jpsi$ and $\gmlg$.
The signal yields in each angular bin were extracted by the global fit
to the corresponding dataset following the aforementioned procedure.
After correction of the extracted signal yields with the detection efficiency,
Fig.~\ref{fig:decay_fit}(a) shows the fit to the
distribution of $|\cos\beta|$ for the sum of
the two dilepton modes; we obtain  $a=0.53\pm0.68$.

\begin{table}[t]
  \scriptsize
   \vspace{-0.25cm}
  \caption{\footnotesize Summary of the systematic uncertainties on the measurement of $\mathcal{B}_{\unit{sig}}$ of $\gm\gm\jpsi$ signal process, $\mathcal{B}_{\chicj}$ for $\chicj$ intermediate processes and the relative branching fractions $R_{MN}$, following the notation convention in Table~\ref{tab:res}. The tot systematic uncertainty is the square root of the sum. A dash (--) means the uncertainty is negligible. Values inside the parentheses are for the $\gm\gm\mm$ mode, while values outside are for the $\gm\gm\ee$ mode. Numbers without brackets represent uncertainties that are common to both modes.}
  \label{tab:sys_err}
  \vspace{-0.3cm}
  \begin{center}
  \begin{tabular}{l|l|ccccccc}
      \hline
      \hline
      \multicolumn{2}{c|}{\footnotesize systematic uncertainty(\%)}
      & ${\footnotesize \mathcal{B}_{\unit{sig}}}$  & {\footnotesize $\mathcal{B}_{\chiczero}$}
      & {\footnotesize $\mathcal{B}_{\chicone}$}    & {\footnotesize $\mathcal{B}_{\chictwo}$}
      & ${\footnotesize R_{01}}$ &  ${\footnotesize R_{02}}$ &  ${\footnotesize R_{21}}$ \bigstrut \\ \hline
      \multicolumn{2}{l|}{\footnotesize lepton track}                     & $2(2)$  & $2(2)$  & $2(2)$  & $2(2)$  \bigstrut \\
      \multicolumn{2}{l|}{\footnotesize photon shower}           & 2  & 2  & 2  & 2  \bigstrut \\
      \multicolumn{2}{l|}{\footnotesize {number of photons} }            & 10(3)  & 1(1)  & 1(1)  & 1(1)   & 2(--)   & 2(--)  & --(--)  \bigstrut \\
      \multicolumn{2}{l|}{\footnotesize KF, $\chisq{KF}$ requirement}           & 2(2)  & 2(2)  & 2(2)  & 2(2)   \bigstrut \\
      \multicolumn{2}{l|}{\footnotesize $\chicj$ widths}            & $^{+15}_{-25}$ & 3  & --  & --  &  4  & 4 &  0.2  \bigstrut  \\
      \multicolumn{2}{l|}{\footnotesize $\rmgmsm$ resolution}                 & 4(5) & --(--)  & --(--)  & --(--)  &   --(--)   &   --(--) &  --(--)\bigstrut  \\
      \multicolumn{2}{l|}{\footnotesize other background}                 & 4(2) & 1(1)  & --(--)  & --(--)  & --(--)  & --(--) & --(--) \bigstrut \\
      \multicolumn{2}{l|}{\footnotesize $\chicj$ interference}           & 5  & 1  & --  & --   & 1  & 1  & -- \bigstrut  \\
      \multicolumn{2}{l|}{\footnotesize fitting}                    & 8(5) & 1(1)  & --(--)  & --(--) & 1(1)  &  1(1)  & --(--) \bigstrut \\
      \multicolumn{2}{l|}{\footnotesize spin-structure}             &  20 & 1 & -- & -- & 1 & 1 & -- \bigstrut  \\
      \multicolumn{2}{l|}{\footnotesize {number of $\psip$ }}       & 4  & 4  & 4 & 4 \bigstrut \\
      \multicolumn{2}{l|}{\footnotesize $\br{{\tiny \jpsi\to\dil}}$}        & 1  & 1  & 1 & 1  \bigstrut \\ \hline
      \multirow{2}{*}{\footnotesize total} & correlated             & 14(8)  &  3(3) & 3(3) &  3(3) & 2(1) & 2(1)
                                 &  --(--)  \bigstrut \\
      &   uncorrelated          & $^{+25}_{-33}$  &  6   & 5 & 5  & 4 & 4 & 0.2 \\    \hline \hline
    \end{tabular}
  \end{center}
\vspace{-0.5cm}
\end{table}

The polarization of $\jpsi$ should be helpful in understanding the
mechanism of the transition process~\cite{Artoisenet:2007xi}. The
polarization parameter $\alpha$ can be evaluated from the angular
distribution of the decay rate, expressed as
$\frac{d\,N}{d\,\cos\theta_\ell} \propto 1+\alpha
\cos^2\theta_\ell$. Here,
$\alpha=\frac{\Gamma_{\unit{T}}-2\Gamma_{\unit{L}}}{\Gamma_{\unit{T}}+2\Gamma_{\unit{L}}}$
(with $\Gamma_{\unit{T}}$ and $\Gamma_{\unit{L}}$ being the
transversely and longitudinally polarized decay widths,
respectively) and the helicity angle $\theta_\ell$ is defined as
the angle of the lepton in the $\jpsi$ rest frame with respect to
the $\jpsi$ boost direction in the laboratory frame. For fully
transverse (longitudinal) polarization, $\alpha=+1(-1)$.
Figure~\ref{fig:decay_fit}(b) shows the distribution of
$|\cos\theta_\ell|$ for the sum of the two dilepton modes,
after correcting the signal yields for the detection efficiency and the lepton final state
radiation effect. Our fit result is $\alpha=0.08\pm0.42$.

Sources of systematic errors on the measurement of branching
fractions are listed in Table~\ref{tab:sys_err}.
Uncertainties associated
with the efficiency of the lepton tracking and identification were
studied with a selected control sample of
$\psip\to\pi^+\pi^-(\dil)_{\jpsi}$.
The potential bias due to limiting the maximum number of photon candidates was studied by varying the limit.
Throughout the photon energy region in this work, detection and energy
resolution of photon are well-modeled within a 1\% uncertainty~\cite{Ablikim:2010zn,Collaboration:2010rc}.
Detector resolution of the $\chicj$ tails is taken into account up to the accuracy of the MC simulation.
The corresponding systematic uncertainty is evaluated by scanning the sizes of smearing parameters within their errors.
For the signal process, the dominant uncertainties are from the description of $\chicj$ line shapes, \eg, $\chicj$ widths.
The sensitivity to the $\chicj$ widths is studied by a comparison of the signal yields based
on different settings of the $\chicj$ widths in modeling the $\chicj$ resonances within the current world-average uncertainties.
Relative changes of the signal detection efficiencies are assigned as 20\%,
by varying the input spin-structure within the measurement uncertainties and
weighting the efficiencies in the Dalitz-like plot of Fig.~\ref{fig:scat}(d).

Many sources of systematic uncertainties in
Table~\ref{tab:sys_err} cancel out when extracting the $\psip$
decay plane parameter $a$ and the $\jpsi$ polarization parameter $\alpha$.
The quadrature sum of the remaining systematic uncertainties are
$^{+0.68}_{-0.27}$ and $^{+0.07}_{-0.14}$ for $a$ and $\alpha$, respectively.

To summarize, the first measurement of the two-photon transition
$\psip\to\gm\gm\jpsi$ was carried out at the BESIII experiment.
The branching fraction is given in Table~\ref{tab:res}, as well
as those of the cascade $E1$ transitions.
The measurement of the two-photon process is consistent with
the upper limit obtained in Ref.~\cite{:2008kb}.
The results for the signal process are presented without considering the
possible interferences between the direct transition and
the $\chicj$ states, due to a lack of theoretical guidance.
The distribution of the normal angle of the
$\psip$ decay plane is characterized by the parameter $a=0.53\pm0.68(\unit{stat})^{+0.68}_{-0.27}(\unit{syst})$,
indicating a preference for a positive value.
The $\jpsi$ polarization parameter $\alpha$ was evaluated as
$0.08\pm0.42(\unit{stat})^{+0.07}_{-0.14}(\unit{syst})$,
demonstrating a competitive mixing of the longitudinal and transverse components.
These results will help constrain the strength of the coupled-channel effect in future theoretical calculation.
The reported branching fractions $\br{\psip\to\gm(\gm\jpsi)_{\chicj}}$
are consistent with the world average results~\cite{pdg2012}.
The reported relative branching fractions of $\mathcal{B}_{\chicj}$ are obtained with the world's best precision.

X.-R. Lu thanks Zhi-Guo He and De-Shan Yang for useful suggestions. We thank the staff of BEPCII and the computing center for their hard efforts. We are grateful for support from our institutes and universities and from the agencies:
Ministry of Science and Technology of China,
National Natural Science Foundation of China,
Chinese Academy of Sciences,
Istituto Nazionale di Fisica Nucleare,
U.S. Department of Energy,
U.S. National Science Foundation,
and National Research Foundation of Korea.


\end{document}